\let\Xdocument\document
\documentclass{iau}
\let\document\Xdocument

\usepackage{amsmath}
\usepackage{graphicx}
\usepackage{multirow}

\begin{document}

\lefttitle{Thomas A Baycroft}
\righttitle{Progress report on the BEBOP search for circumbinary planets with radial velocities}

\jnlPage{1}{7}
\jnlDoiYr{2021}
\doival{10.1017/xxxxx}
\volno{382}

\aopheadtitle{Proceedings Kavli-IAU Symposium}
\editors{A. Lemaitre \&  A.-S. Libert, eds.}

\title{Progress report on the BEBOP search for circumbinary planets with radial velocities}

\author{Thomas A. Baycroft$^1$, Amaury H.M.J. Triaud$^1$, and Dong Lai$^2$}
\affiliation{$^1$University of Birmingham, Birmingham, United Kingdom\\  $^2$ Cornell University, Ithaca, NY, United States of America}

\begin{abstract}
The BEBOP (Binaries Escorted By Orbiting Planets) survey is a search for circumbinary planets using the radial velocity spectrographs HARPS and SOPHIE, currently focusing on single-lined binaries with a mass ratio \(<\) 0.3. Circumbinary systems are an important testing ground for planet formation theories as the dynamically complex influence of the binary makes planet formation and survival more difficult. Here we present the results of the survey so far including: confirmed planets such as BEBOP-1c the first circumbinary planet detected in radial velocity; the status of our observations; and preliminary occurrence rates. We compare the early results of the radial velocity survey to the population of circumbinary planets discovered in transit, and suggest that there may be a population of inflated planets close to the inner binary which are detectable in transit but more difficult in radial velocity. Using time-lag tidal theory, we show that this inflation is unlikely caused by tides.
\end{abstract}


\maketitle

\section{Introduction}
Over 5000 planets orbiting other stars have been confirmed to date. The majority of these have been detected around single stars. This is widely assumed to be an observation bias rather a reflection of the underlying population.

Planets in binary systems come in two flavours: planetary-type (p-type), in a wider orbit around the centre of mass of the two stars; and satellite-type (s-type), orbiting in a close orbit around one of the two stars. The planetary-type are also known as circumbinary planets, and these are the rarer of the two flavours with only 15 such planets known to orbit main-sequence binaries \citep{standing_radial-velocity_2023}. Again, this is most likely due to lack of observations rather than a true dearth of such planets. 

Around the inner binary, there is an instability zone, in which planets would not be dynamically stable. The extent of this zone has been numerically determined \citep[e.g.][]{dvorak_stability_1989,holman_long-term_1999}, in general it is between \(4\times\) and \(6\times\) the period of the inner binary. Around the inner binary there is also a zone in which, according to current theories, planet formation is inhibited \citep{moriwaki_planetesimal_2004,paardekooper_how_2012}, this extends to \(\sim 50 \times\) the binary period. This zone is much more extended than the instability zone, meaning that a planet can live on an orbit in a region where it cannot have been born. Hence, circumbinary planets that are found to be near the instability zone must have formed further out in the protoplanetary disc and migrated inwards to their current location \citep{kley_modeling_2014}. Circumbinary planets are therefore a great probe of migration pathways as they can have populations that are \textit{known} to have migrated, which cannot be said for most exoplanets hosted by single stars.

Circumbinary exoplanets have been detected in transit by both the \textit{Kepler} \citep[e.g.][]{doyle_kepler-16_2011} and \textit{TESS} \citep[e.g.][]{kostov_toi-1338_2020} space telescopes. There are 14 of them in 12 systems. Their orbital periods put almost all of them very close to the instability region, and therefore means they most likely have had to migrate to reach their current locations. The fact that they are all so close to the instability limit has been interpreted as a sign that there is a pile-up of circumbinary planets near the instability limit at around \(6\times\) the binary period, and that this is not due to an observational bias \citep{martin_planets_2014}.

The radial-velocity method discovered planets around single stars before the transit method. For circumbinary planets it is the other way around, as the binary itself has hampered efforts to detect them. Double-lined binaries \citep[which account for 92\% of binaries;][]{kovaleva_visual_2016} are those where spectral lines from both components' are combined into the observed spectrum. The spectral lines' interference with each other in the standard cross-correlation technique for radial-velocity extraction adds extra scatter to the data which makes it difficult to detect planets. One way to circumvent this is to observe single-lined binaries. With a mass ratio \(<0.3\), the secondary star in these cases is not bright enough to cause a significant effect on the primary star's measured radial velocities. This is the method that has been used by the BEBOP survey whose results so far are presented here.

\section{The BEBOP survey and published results}
The BEBOP (Binaries Escorted By Orbiting Planets) survey, is a blind, radial velocity survey on a sample of single-lined eclipsing binaries \citep{martin_bebop_2019}. These were identified from the EBLM \citep[Eclipsing Binaries with Low Mass companions;][]{triaud_eblm_2013} survey and from \textit{TESS}. Data has been collected with the HARPS spectrograph \citep{mayor_setting_2003} and the SOPHIE spectrograph \citep{perruchot_sophie_2008} for up to 5 years. There is a main sample of around 100 stars, and a secondary sample also numbering around 100.

The first detection from the BEBOP survey was the confirmation of Kepler-16\,b \citep{triaud_bebop_2022}. This had been the first circumbinary planet discovered in transit \citep{doyle_kepler-16_2011}, and BEBOP independently detected it using the radial velocity method. This provided the proof-of-concept that the detection of circumbinary planets in radial velocity is attainable, as well as confirmed the mass of the planet which had been determined from eclipse timing variations.

The second detected planet from the BEBOP survey is a new discovery, BEBOP-1/TOI-1338\,c \citep{standing_radial-velocity_2023}. This is a multi-planet system with an inner transiting circumbinary planet discovered by \textit{TESS} \citep{kostov_toi-1338_2020} and an outer planet discovered in radial velocity by BEBOP \citep{standing_radial-velocity_2023}. The outer planet has not yet been seen in transit, this is likely due to the longer period as well as the fact that circumbinary planets alternate being in and out of \textit{transitability} \citep{martin_circumbinary_2015}. The inner planet is not yet detected with the radial velocity data, this allows us to put an upper limit on the mass of planet b of \( 21.8\pm0.9 \mathrm{\,M_{\oplus}}\). This along with the planetary radius being \(6.85\pm0.19 \mathrm{\,R_{\oplus}}\) \citep{kostov_toi-1338_2020} implies a bulk density \(\rho_b \lesssim 0.36 \mathrm{\,g\,cm^{-3}}\). This means that planet b is under-dense, perhaps meaning that it is inflated in some way.

\section{Results and Occurrence rates}

We now present very preliminary occurrence rates of the BEBOP survey based on detected and candidate planets. 

The radial velocity analysis undertaken by BEBOP uses the analysis package {\tt kima} \citep{faria_kima_2018}, this utilises diffusive nested sampling \citep[DNEST4;][]{brewer_dnest4_2018} to efficiently explore parameter space. The trans-dimensional nested sampling allows to explore different numbers of planetary signals in the data and to compute a Bayes' Factor comparing them. We claim strong evidence for a planet when the Bayes' Factor for the model with a planet to the one without is greater than 150 \citep[as in][]{kass_bayes_1995}. {\tt kima} has a BINARIES radial-velocity model tailored to fitting binary orbits by including apsidal precession of the binary and accounting for GR and tidal effects \citep{baycroft_improving_2023}. The binary orbit can also be fit with independent priors to the planets meaning this is very suited to the detection of circumbinary planets in radial velocity.

From the most recent analysis of the BEBOP sample we have three categories of \textit{planets}: detected, candidate, long-period candidate. The detected planets consist of BEBOP-1c as well as two currently unpublished planets with Bayes' Factors exceeding 150 putting them in the confident detection band. There are 13 candidate planets, these have a Bayes' Factor less than 150 so are not (yet) formally classed as detections. The long-period candidates consists of signals which have a period longer than the baseline of observations, these may have Bayes Factors exceeding 150 but the nature and parameters of the signal are uncertain. There are 9 of these signals for which we take a lower-bound on the orbital period and mass based on the shortest period circular orbit within the \(1\sigma\) confidence interval for the orbital period.

\begin{figure}
    \centering
    \includegraphics[scale=.8]{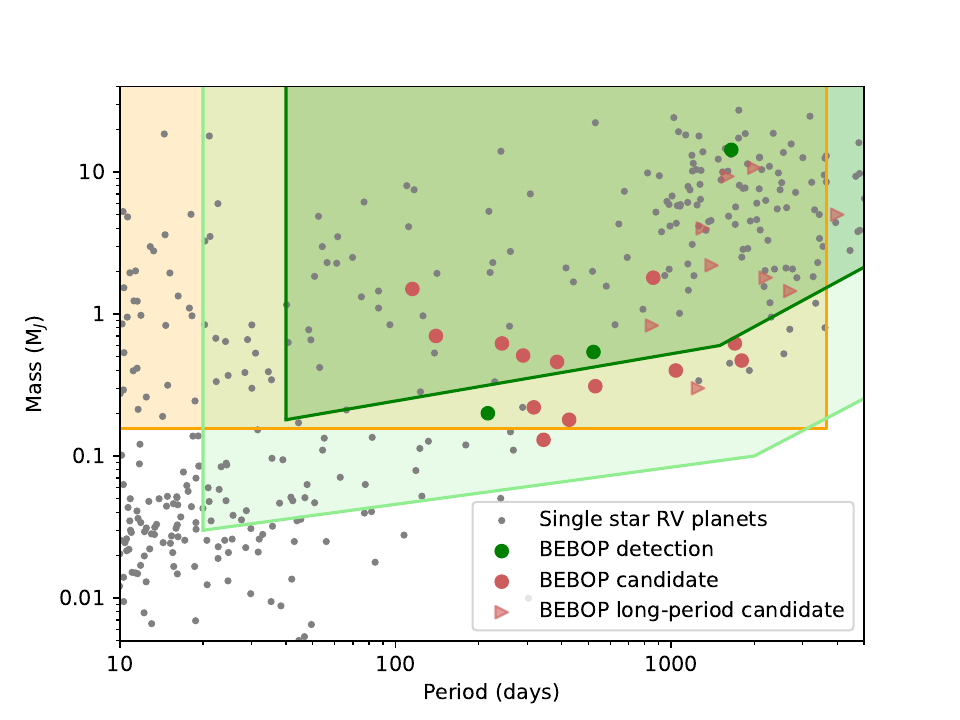}
    \caption{Preliminary occurrence plot showing the BEBOP planets and candidates in colour with the radial velocity population of planets around single stars show in grey (https://exoplanetarchive.ipac.caltech.edu/). Dark green the median sensitivity of BEBOP. Light green the best sensitivity of BEBOP. Yellow the area inside which
    \citet{mayor_harps_2011} find an occurrence rate of 14-15\(\%\).}
    \label{fig:occurences}
\end{figure}

The 3 detections, 13 candidates and 9 long-period candidates are shown on a mass-period plot in Figure \ref{fig:occurences} with the colours and shapes of the points differentiating between detections, candidates and long-period candidates. The green areas show the best and median sensitivity of the BEBOP survey, the grey points are radial velocity planets around single stars. The yellow area is from \citet{mayor_harps_2011}, they find that 14-15\% of single stars have a planet in this yellow area. We can compare this to the BEBOP planets to see that currently the circumbinary planet population is consistent with this. From a simplistic occurrence rate we obtain between 3\% (if only counting the detections) and 23\% (if also counting the candidates) of binaries host an exoplanet in the yellow box. These numbers do not take into account the completeness of the survey, and this will be done more carefully once the survey has been completed.

\section{Comparison to transiting population}

\begin{figure}
    \centering
    \includegraphics[scale=.263]{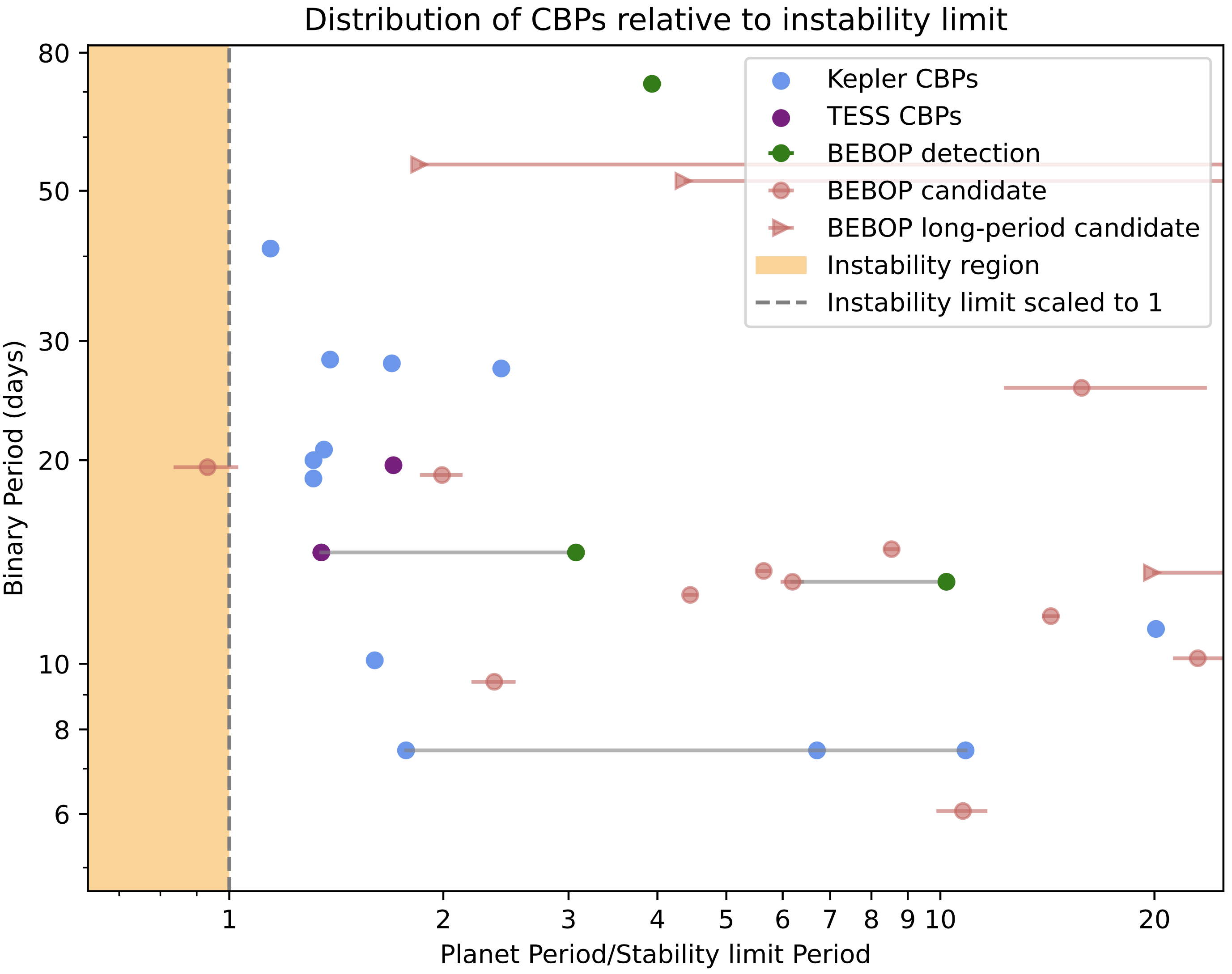}\\
    \includegraphics[scale=.30]{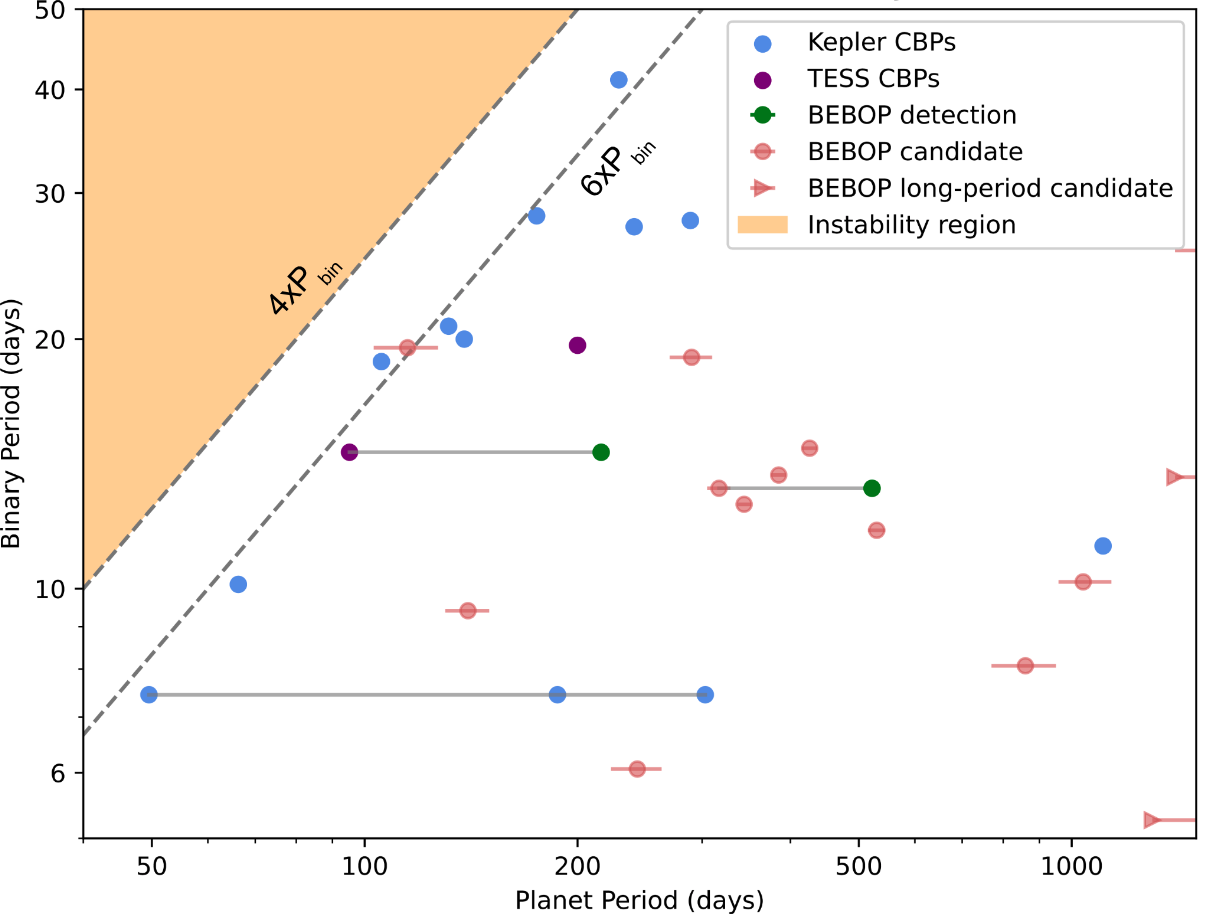}
    \caption{Period distribution of circumbinary planets detected in transit and the BEBOP radial velocity planets and candidates. Top: The periods are scaled by the period that a planet would have if on a circular orbit at the stability boundary. Bottom: The true orbital periods of the planets are shown, with the dashed lines showing four times and six times the binary period.}
    \label{fig:periods}
\end{figure}

Figure \ref{fig:periods} shows both the transiting population of circumbinary planets as well as the detections, candidates and some of the long-period candidates from BEBOP. The transiting population shows a pile-up near the instability limit, this pile-up is not seen in the radial velocity population. The lower plot motivates why this is not solely due to the transit method's geometric observation bias towards short-period orbits. This bias is solely dependent on the distance of the planet from the star, this means that a vertical strip in the bottom figure should show planets that are all equally as likely to transit \citep{martin_planets_2014}. The fact that all the transiting planets are near the \textit{top} of such strips implies that this pattern cannot solely be an observation bias due to the distance of such planets from the central star.

To explain why the pile-up is seen with transiting planets, but not with radial-velocity planets, a possible solution is that circumbinary planets near the instability limit are inflated, making them both easier to detect in transit (increasing their radius), and harder with radial-velocity (reducing their mass). One example of this is TOI-1338/BEBOP-1 b which is inflated as described above.

\section{Tidal energy dissipation}

One possible mechanism for the inflation of circumbinary planets close to the stability limit is from tidal energy dissipation. From the any of the two stars' reference frames, the planet occupies a constantly  eccentric orbit. Tides raised on the planet will attempt to circularise its orbit and synchronise its rotation \citep[e.g.][]{Guillot1996}. However, the planet's motion about the binary already being circular (or close to it), tidal equilibrium is prevented, and therefore more energy could be dissipated over the lifetime of the system. We perform a back of the envelope calculation for TOI-1338/BEBOP-1\,b using time-lag tidal theory \citep{Hut1981}. 

We consider a tide raised on the planet of mass \(m_{\rm p}\) by the secondary of mass \(m_2\,(\lesssim m_1)\). Relative to the barycentre, the position of the planet is \(\vec{r_{\rm p}}(t)\) and that of the secondary is \(\vec{r_2}(t)\), and we write \(\vec{d_{\rm p}}(t) = \vec{r_{\rm p}}(t) - \vec{r_2}(t)\). Now \(d_{\rm p}(t)\) varies with mean frequency \(\Omega_2 - \Omega_{\rm p} = \Delta\Omega\), with \(\Omega_{\rm p}\) and \(\Omega_2\) the orbital angular velocities of the planet and secondary. We can give an approximate expression for the tidal kinetic energy: 
\begin{equation}
    E_{\rm K} \sim k_2\frac{m_2^2}{m_{\rm p}}\frac{R_{\rm p}^8}{d_{\rm p}^6}(\Delta\Omega)^2,
\end{equation}
where \(k_2\) is the Love number and \(R_{\rm p}\) is the planet's radius. Then the tidal energy dissipation can be taken as \(\dot{E}_{\rm K} \sim \frac{E_{\rm K}}{t_{\rm dissipation}} \sim E_{\rm K}\frac{\Delta\Omega}{Q}\) where \(Q\) is the tidal quality factor. This leads us to an expression for the tidal energy dissipation rate:
\begin{equation}
    \dot{E}_{\rm K} \sim \frac{k_2}{Q}m_{\rm p}\left(\frac{m_2}{m_{\rm p}}\right)^2\left(\frac{R_{\rm p}}{d_{\rm p}}\right)^6R_{\rm p}^2(\Delta\Omega)^3.
\end{equation}

Substituting in parameters for the TOI-1338/BEBOP-1 system, and assuming \(\frac{k_2}{Q} \sim 10^{-5}\), yields an order of magnitude estimate \(\dot{E}_{\rm K} \sim 10^{21} \mathrm{\,erg/s}\). An order-of-magnitude estimate for the energy coming from insolation gives \(L_{\rm in} \sim 10^{28}\mathrm{\,erg/s}\). This implies that energy dissipation from tides is not significant enough to have an impact on the planetary radius, meaning there is likely another explanation.

However, this is still an open question, more circumbinary planets detected with both radial-velocities and with transits will help understand the populations better, and more theoretical models of circumbinary planets, including but not limited to models of the tidal effects, will be beneficial.

\section{Conclusions}

We have presented the BEBOP radial velocity survey for circumbinary planets orbiting single-lined spectroscopic binaries, as well as preliminary results from the survey. 

We have compared the detections and candidates to both the population of radial velocity planets orbiting single stars and the transiting circumbinary planet population. The current results do not show a significant difference in occurrence rate from the single star population.

The pile-up of planets seen in the transiting population is not as apparent in the radial velocity population. We suggest that a possible explanation for this could be that the planets that migrate close to the instability limit are inflated. We perform a back-of-the-envelope calculation that casts doubt on this inflation being caused by tides raised by the binary.

\bibliographystyle{iaulike}
\bibliography{CPSII_proceedings}

\end{document}